\documentclass[12pt,preprint,onecolumn]{emulateapj}
\usepackage{latexsym}
\usepackage{amssymb}
\usepackage{amsmath}
\newcommand{\etal}{{et al.\ }}

\newcommand{\beq}{\begin{equation}}
\newcommand{\eeq}{\end{equation}}
\newcommand{\ba}{\begin{eqnarray}}
\newcommand{\ea}{\end{eqnarray}}
\def\spose#1{\hbox to 0pt{#1\hss}}
\newcommand{\lta}{\mathrel{\spose{\lower 3pt\hbox{$\mathchar"218$}}
      \raise 2.0pt\hbox{$\mathchar"13C$}}}
\newcommand{\gta}{\mathrel{\spose{\lower 3pt\hbox{$\mathchar"218$}}
      \raise 2.0pt\hbox{$\mathchar"13E$}}}
\def\HI{\hbox{H~$\scriptstyle\rm I$}}

\def\nHI{{\rm HI}}

\def\nHeI{{\rm HeI}}
\def\nHeII{{\rm HeII}}

\def\HeI{\hbox{He~$\scriptstyle\rm I$}}
\def\HeII{\hbox{He~$\scriptstyle\rm II$}}

\def\HeIII{\hbox{He~$\scriptstyle\rm III$}}
\def\CIV{\hbox{C~$\scriptstyle\rm IV$}}
\def\CIII{\hbox{C~$\scriptstyle\rm III$}}
\def\CII{\hbox{C~$\scriptstyle\rm II$}}

\def\OVI{\hbox{O~$\scriptstyle\rm VI$}}
\def\OII{\hbox{O~$\scriptstyle\rm II$}}
\def\OIII{\hbox{O~$\scriptstyle\rm III$}}
\def\SiII{\hbox{Si~$\scriptstyle\rm II$}}
\def\SiIII{\hbox{Si~$\scriptstyle\rm III$}}
\def\SiIV{\hbox{Si~$\scriptstyle\rm IV$}}
\def\cmm{\,{\rm cm^{-2}}}

\def\uvunits{\,{\rm ergs\,cm^{-2}\,s^{-1}\,Hz^{-1}\,sr^{-1}}}
\def\emunits{\,{\rm ergs\,s^{-1}\,Hz^{-1}\,Mpc^{-3}}}

\def\Lya{Ly$\alpha$}
\def\Lyb{Ly$\beta$}
\def\Lyg{Ly$\gamma$}
\lefthead{Madau \& Haardt}
\righthead{Sawtooth spectrum of the far-UV background}

\begin{document}
\submitted{ApJL, in press}
\title{He II absorption and the sawtooth spectrum of the cosmic far-UV background}
 
\author{Piero Madau\altaffilmark{1} \& Francesco Haardt\altaffilmark{2}}

\altaffiltext{1}{Department of Astronomy \& Astrophysics, University of 
California, Santa Cruz, CA 95064.}
\altaffiltext{2}{Dipartimento di Fisica e Matematica, Universit\'a dell'Insubria, Via Valleggio 
11, 22100 Como, Italy.}

\begin{abstract}
Cosmic ultraviolet background radiation between 3 and 4 Ryd is reprocessed by resonant
line absorption in the Lyman series of intergalactic \HeII. This process results 
in a sawtooth modulation of the radiation spectrum from the \HeII\ \Lya\ frequency 
to the Lyman limit. The size of this modulation is a sensitive probe of the epoch of helium 
reionization and of the sources that keep the intergalactic medium (IGM) highly ionized. For large absorption 
opacities, the background intensity will peak at frequencies just above each resonance, go to zero 
at resonance, and fluctuate greatly just below resonance. The \HeII\ sawtooth modulation may be one of 
the missing ingredients needed in the modelling of the abundances of metal ions like \CIII\ and \SiIV\ 
observed in the IGM at redshift 3. 
\end{abstract}


\section{Introduction}

The intensity and spectrum of the cosmic ultraviolet background are one of the most 
uncertain yet critically important astrophysical input parameters for cosmological 
simulations of the intergalactic medium (IGM) and early reionization. Theoretical 
models of such diffuse radiation field can help interpret quasar absorption-line 
data and derive information on the distribution of primordial baryons 
(traced by \HI, \HeI, \HeII\ transitions) and of the nucleosynthetic products of 
star formation (\CIII, \CIV, \SiIII, \SiIV, \OVI, etc.). Because of the high 
ionization threshold (54.4 eV) and small photoionization cross-section 
of \HeII, and of the rapid recombination rate of \HeIII, the double ionization of 
helium is expected to be completed by hard UV-emitting quasars around the peak of their 
activity at $z\approx 3$ (e.g. Madau \etal 1999; Sokasian \etal 2002), much later than the 
reionization of intergalactic \HI\ and \HeI. Several observations have provided controversial
evidence for a transition in some properties of the IGM around the predicted epoch of helium 
reionization, from the \HeII\ Gunn-Peterson trough measured in the spectra of $z>2.8$ quasars
(e.g. Jakobsen \etal 1994; Reimers \etal 1997; Heap \etal 2000; Smette \etal 2002), to the 
possible detection of an increase in the temperature of the IGM (e.g. Ricotti \etal 2000; 
Schaye \etal 2000; McDonald \etal 2001; Zaldarriaga \etal 2001;
Theuns \etal 2002; Bernardi \etal 2003), to the claimed sharp evolution in the 
column density ratios of metal line absorbers (Songaila \etal 1998; Boksenberg 
\etal 2003; Agafonova \etal 2007). 

With the imminent installation of the Cosmic Origins Spectrograph on board of the 
{\it Hubble Space Telescope}, the quantity and quality of far-UV observations of 
the IGM will improve significantly. Numerical simulations of patchy \HeII\ reionization 
(Paschos \etal 2007; McQuinn \etal 2008) are already shedding new light on the 
nature of such late reheating process and its potential impact on observables.  
In this {\it Letter} we return to the theory of cosmological radiative transfer and 
to the atomic processes that shape the spectrum of the far-UV        
background. We address a hitherto unnoticed effect, resonant absorption by the \HeII\ Lyman 
series, and show that this process will produce a 
sawtooth modulation of the radiation spectrum between 3 and 4 Ryd. The size of this modulation  
depends sensitively on the abundance of \HeII\ in the IGM, and may in turn be a crucial 
factor in determining the abundance of metal ions like \CIII, \SiIII, and \SiIV\ in the IGM.    
The analogous modulation between 0.75 and 1 Ryd from hydrogen
line absorption was first studied by Haiman \etal (1997) in the limiting case of a fully
neutral IGM.

\section{Spectral filtering by the IGM}

We treat the radiation field $J_\nu(z)$ as a uniform, isotropic background, and
include the reprocessing of UV radiation in a clumpy IGM (HM96). The specific intensity 
(in $\uvunits$) at redshift $z_o$ and observed frequency $\nu_o$ is given by 
\begin{equation}
J_{\nu_o}(z_o)={c\over 4\pi}\int_{z_o}^{\infty}\, {dt\over dz}dz 
{(1+z_o)^3 \over (1+z)^3} \epsilon_\nu(z) e^{-\bar\tau}, 
\label{Jnu}
\end{equation}
where $\nu=\nu_o(1+z)/(1+z_o)$, $(dt/dz)=[H(z)(1+z)]^{-1}$, $H(z)$ is the Hubble 
parameter, $e^{-\bar\tau}\equiv \langle e^{-\tau}\rangle$ is the average 
cosmic transmission over all lines of sight, and $\epsilon_\nu$ (in $\emunits$) is 
the proper volume emissivity. The effective continuum (LyC) optical depth 
between $z_o$ and $z$ from Poisson-distributed absorbers is 
\begin{equation}
\bar\tau(\nu_o)\equiv \bar\tau_c=\int_{z_o}^z\,
dz'\int_0^{\infty}\, dN_\nHI\, f(N_\nHI,z') (1-e^{-\tau_c}), \label{tau}
\end{equation}
where $f(N_\nHI,z')$ is the bivariate distribution of absorbers in redshift and 
column density along the line of sight, and $\tau_c$ is the LyC optical depth through 
an individual cloud of hydrogen and helium column densities $N_\nHI$, $N_\nHeI$, 
and $N_\nHeII$. The effective line absorption optical depth is instead 
\begin{equation}
\bar\tau_n(z)={(1+z)\nu_n\over c} \int dN_\nHI\,f(N_\nHI,z)W_n,
\label{taui}
\end{equation}
where $\nu_n$ is the frequency of the $1s \rightarrow np$ Lyman series transition ($n>2$) and $W_n$ 
is the rest equivalent width of the line expressed in wavelength units. 

\subsection{Resonant \HeII\ absorption}

The far-UV metagalactic flux has long been known to be partially
suppressed by the \HeII\ and \HI\ continuum opacity of the IGM
(e.g. Miralda-Escude \& Ostriker 1990; Madau 1992), but little attention has been 
given to \HeII\ line absorption. Photons passing through the \HeII\ \Lya\ resonance 
are scattered until 
they redshift out of resonance, without any net absorption: aside from the photoionization 
of \HI\ and \HeI, the only other \HeII\ \Lya\ destruction mechanism, two-photon decay, 
is unimportant in the low density IGM at the redshifts of interest.\footnote{Another 
conversion process of \HeII\ \Lya\ photons, the \OIII\ Bowen fluorence mechanism
(Kallman \& McCray 1980), can be neglected in nearly primordial intergalactic gas.}~
This is not true, however, for photons passing through a \HeII\ Lyman series resonance between 
the Lyman limit at energy $h\nu_L=4$ Ryd and the \HeII\ \Lyb\ at $h\nu_\beta=3.56$ Ryd.
If the opacity of the IGM in the Lyman series lines is large, \Lyb\ and higher Lyman 
line photons will be absorbed and degraded via a radiative cascade rather than escaping 
by redshifting across the line width. The net result is a sawtooth modulation
of the spectrum between 3 and 4 Ryd, and a large discontinuous step at the \HeII\ \Lya\
frequency, as we show below.

Consider for example radiation observed at frequency $\nu_o<\nu_\beta$ and redshift $z_o$.
The resonant absorption cross-section is a narrow, strongly-peaked function, 
different lines dominate the opacity at different absorption redshifts,
and the line and continuum transmission can be treated as independent 
random variables (e.g. Madau 1995). Photons emitted between $z_o$ and 
$z_\beta=(1+z_o)(\nu_\beta/\nu_o)-1$ can reach the observer without undergoing 
resonant absorption. Photons emitted between $z_\beta$ and $z_\gamma=(1+z_o)
(\nu_\gamma/\nu_o)-1$ pass instead through the \HeII\ \Lyb\ resonance at $z_\beta$ and are 
absorbed. Photons emitted between $z_\gamma$ and $z_\delta=(1+z_o)(\nu_\delta/\nu_o)-1$ 
pass through both the \HeII\ \Lyb\ and the \Lyg\ resonances before reaching
the observer. The background intensity can then be written as 
\begin{equation}
J_{\nu_o}(z_o)=I(z_o,z_\beta)+I(z_\beta,z_\gamma)e^{-\bar\tau_\beta}+
I(z_\gamma,z_\delta)e^{-\bar\tau_\beta-\bar\tau_\gamma}+....
+I(z_L,\infty)e^{-\sum_n\bar\tau_n},
\label{Jbeta}
\end{equation}
where we denote with the symbol $I(z_i,z_j)$ the right hand side of equation ({\ref{Jnu}) 
integrated between $z_i$ and $z_j$ with $\bar\tau \equiv \bar\tau_c$. Here $z_L=(1+z_o)(\nu_L/\nu_o)-1$, 
the LyC opacity $\bar\tau_c$ in all $I$ integrals except the last 
(where \HeII\ must be added) includes only \HI\ and \HeI\ absorption, 
and $\bar\tau_\beta, \bar\tau_\gamma, \bar\tau_\delta,...$ are the \HeII\ Lyman 
series effective opacities at redshift $z_\beta, z_\gamma, z_\delta,...$.   
Equation (\ref{Jbeta}) is easily generalized to higher frequencies,
e.g. for $\nu_\beta<\nu_o<\nu_\gamma$ the first two terms must be replaced by the integral 
$I(z_o,z_\gamma)$. 

\begin{figure}[thb]
\centering
\includegraphics*[width=0.48\textwidth]{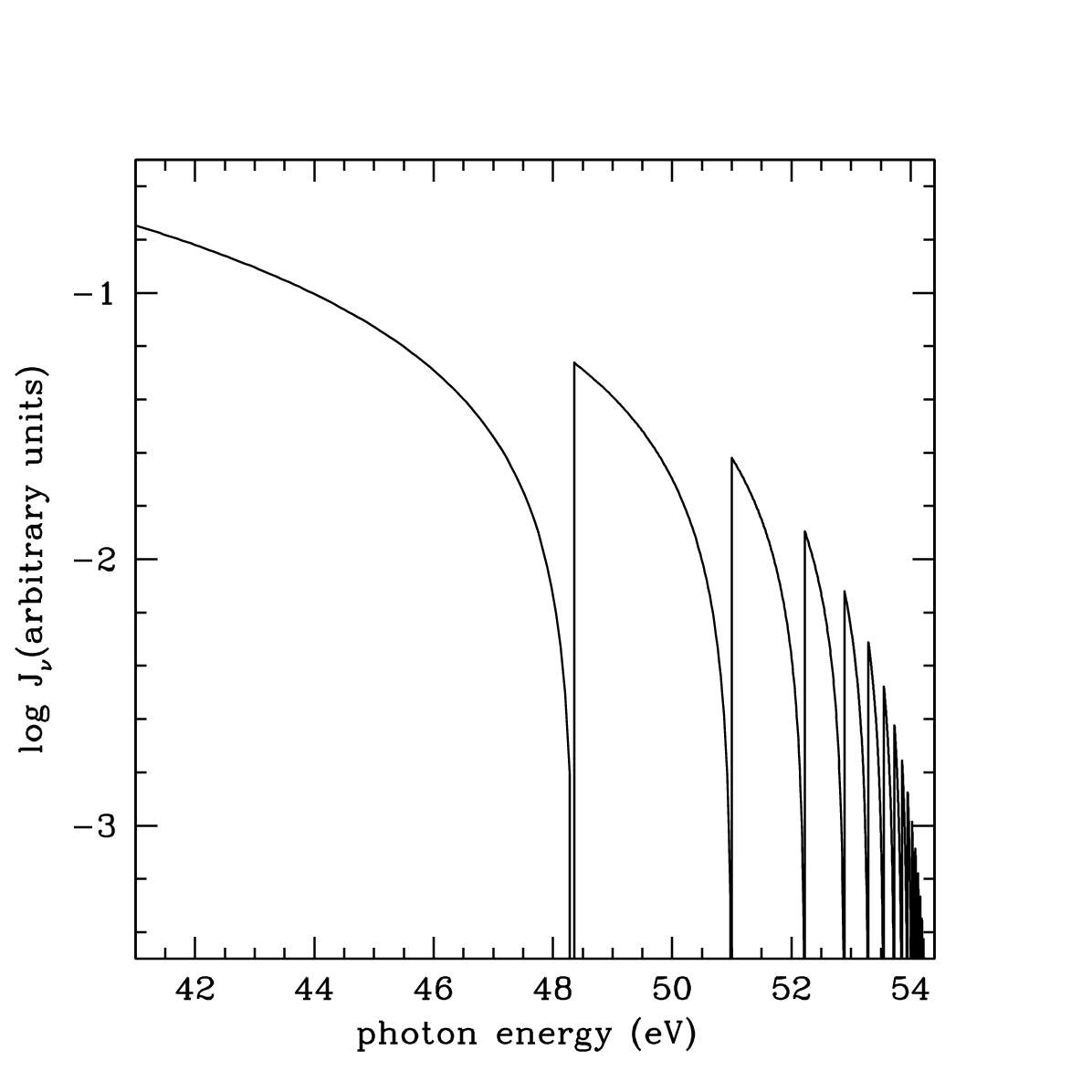}
\includegraphics*[width=0.48\textwidth]{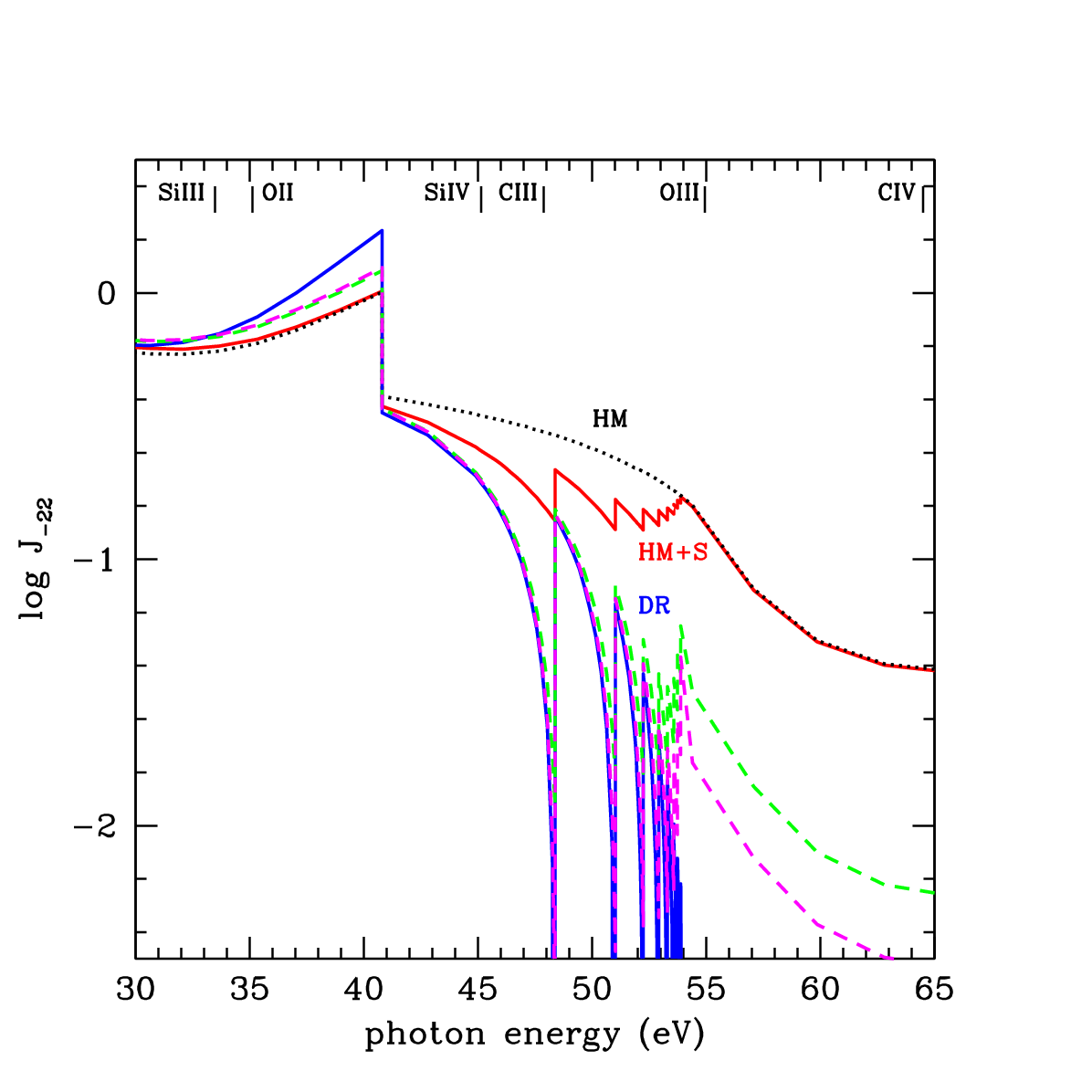}
\caption{{\it Left:} The sawtooth modulation of the metagalactic flux between 3 and 4 Ryd
produced by resonant absorption in the Lyman series of intergalactic \HeII. The background specific intensity
was computed assuming negligible continuum absorption, large line opacities, an integrand in 
equation (\ref{Jnu}) independent of redshift, and a proper emissivity $\epsilon_\nu\propto \nu^0=$const. 
{\it Right:} The far-UV background intensity (in units of $10^{-22}\,\uvunits$)
at redshift 3 from quasar sources, computed using the radiative
transfer code CUBA. Several models are compared:   
1) ``HM", where resonant absorption from the Lyman series of singly-ionized helium was neglected ({\it dotted
line}); 
2) ``HM+S", where the sawtooth modulation of \HeII\ was added ({\it solid red line}). Both these models assume 
photoionization equilibrium with a uniform radiation field following HM96, and yield a value 
\HeII/\HI=35 in optically thin absorbers; and 
3) ``DR" (for ``delayed reionization"), where the \HeII/\HI\ ratios were artificially increased 
to 160 ({\it dashed green line}), 250 ({\it dashed magenta line}), and 530 ({\it solid blue line}).
All 5 models assume the quasar emissivity and absorbers distribution given in eqs. (\ref{epsi}) 
and (\ref{fN}) of the text. The positions of the ionization thresholds of different ions 
are indicated by tick marks. 
}
\label{fig1}
\end{figure}

The effect of the sawtooth modulation is best depicted in the idealized case of negligible continuum
absorption, $\bar\tau_c\rightarrow 0$, and large line opacity, $\bar\tau_n\rightarrow \infty$
(this is similar to the hydrogen case studied by Haiman \etal 1997). The 
ensuing radiation flux is shown in Figure \ref{fig1}, where we have also assumed for 
simplicity that the integrand in equation (\ref{Jnu}) is independent of redshift and 
the proper emissivity is $\epsilon_\nu\propto \nu^0=$const. Note how only sources 
between the observer and the ``screen'' redshift $z_n=(1+z_o)(\nu_n/\nu_o)-1$ corresponding to
the frequency of the nearest Lyman series line above $\nu_o$ are not blocked from view:
the background flux peaks at frequencies just above each resonance, as the first 
integral in equation (\ref{Jbeta}) extends over the largest redshift path, and 
goes to zero only at resonance.

\subsection{\HeII\ \Lya\ riemission}

The usual assumption that each photon entering a Lyman series resonance causes a radiative
cascade that terminates in a \Lya\ photon requires full $l$-mixing of the $2s-2p$
levels (Seaton 1959). Collisions are infrequent in the low-density IGM, however, and most radiative
cascades from an $np$ state terminate instead in two-photon $2s \rightarrow 1s$
emission (Hirata 2006). The fraction, $f_n$, of decays that generates \Lya\ photons can be
determined from the selection rules and the decay probabilities, and it is 
$f_n=(1,0,0.2609,0.3078,0.3259...)$ for $n=(2,3,4,5,6...)$ (Pritchard \& Furlanetto 2006).
Without $l$-mixing, the quantum selection rules forbid a \Lyb\ photons from being converted 
into \Lya, while at large $n$ the
conversion fraction asymptotes to 0.36. Let now $J_{\nu_\gamma}(z_\alpha)$ 
be the background intensity measured just above the \HeII\ \Lyg\ resonance at redshift 
$z_\alpha=(1+z_o)(\nu_\alpha/\nu_o)-1$. The flux that is absorbed and converted into \Lya\ is then 
$f_4\times J_{\nu_\gamma}(z_\alpha)[1-e^{-\bar\tau_\gamma(z_\alpha)}]$. The additional flux observed at 
frequency $\nu_o\le \nu_\alpha$ and redshift $z_o$ from this process is then
\begin{equation}
\Delta J_{\nu_o}(z_o)=\left(\frac{\nu_o}{\nu_\alpha}\right)^3\,e^{-\bar\tau_c(\nu_o,z_o,z_\alpha)}\,
[f_4\times J_{\nu_\gamma}(z_\alpha)(1-e^{-\bar\tau_\gamma(z_\alpha)})].
\label{Jalpha}
\end{equation}
When summing up over all Lyman series lines, the term 
in square brackets must be replaced by $\sum_{n>3}[f_n\times J_{\nu_n}(z_\alpha)(1-e^{-\bar\tau_n
(z_\alpha)})]$.\footnote{Note that the analogous equation (7) of Haiman \etal (1997) 
for the reprocessing of hydrogen Lyman series radiation erroneously includes a term 
$(\nu_n/\nu_\alpha)$ to account for the conversion of higher energy photons into \Lya. 
This factor is spurious since in the process the specific intensity is conserved.}

\section{UV metagalactic flux}   

We now compute the integrated far-UV background from quasar sources at redshift 3, including the 
effect of \HeII\ resonant absorption.  We parameterize the recent determination of the {\it comoving} 
quasar emissivity at 1 Ryd by Hopkins \etal (2007) as  
\begin{equation}
\varepsilon_{\rm ion}(z)=(4\times 10^{24}\,\emunits)\,(1+z)^{4.68}\,{e^{-0.28z}\over e^{1.77z}+26.3}.
\label{epsi}
\end{equation}
At $z=3$, this yields a value that is 1.4 times lower than that used in Madau \etal (1999). Assuming a 
power-law far-UV spectrum with spectral index $-1.57$ (Telfer \etal 2002), the adopted {\it proper} 
emissivity becomes $\epsilon_\nu(z)=(1+z)^3\,\varepsilon_{\rm ion}(z)\,(h\nu/1\,{\rm Ryd})^{-1.57}$.
To compute the effective opacity of the IGM we use the standard parameterization for the distribution 
of absorbers along the line of sight,
\begin{equation}
f(N_\nHI,z)=A\,N_\nHI^{-\beta}\,(1+z)^\gamma,
\label{fN}
\end{equation}
with $(A,\beta,\gamma)=(1.4,1.5,2.9)$ over the column density range $10^{11}<N_\nHI<10^{17.2}\,\cmm$, 
and $(A,\beta,\gamma)=(5,1.5,1.5)$ for $N_\nHI>10^{17.2}\,\cmm$ (e.g. Kim \etal 1997; Hu \etal 
1995; Meiksin \& Madau 1993; Petitjean \etal 1993; Tytler 1987).  
Here, the normalization $A$ is expressed in units of $10^7\,{\rm cm}^{-2(\beta-1)}$. 
The high-column density distribution agrees with the results of Stengler-Larrea \etal (1995),
while the low-column density distribution produces an \HI\ \Lya\ effective opacity 
at $z=3$ of 0.41 as in Faucher-Giguere \etal (2008). The detailed redshift evolution of 
the quasar emissivity and IGM opacity are not important for the problem at hand, since the 
measured -- at 1 Ryd -- and inferred -- at 4 Ryd -- absorption distances for ionizing radiation
at redshift 3 are quite small, and the background flux in the relevant energy range is 
largely determined by local sources. 

We have used the cosmological radiative transfer code CUBA to follow the propagation 
of UV radiation through a partially ionized inhomogeneous medium (HM96). The code uses a multi-zoned
approximation to model the physical conditions within absorbing systems and infer the amount of 
singly-ionized helium that is present along the line of sight from the well-measured $N_\nHI$ distribution.
We include the reprocessing into \HeII\ \Lya\ and two-photon continumm from resonant 
absorption in the Lyman series (as detailed
in the preceding section) as well as continuum absorption, and assume photoionization equilibrium with
a uniform radiation field. The resulting far-UV background intensity at redshift 3 is shown 
in Figure \ref{fig1} for model ``HM", where resonant absorption from the Lyman series of \HeII\ is neglected, 
and model ``HM+S", where the sawtooth modulation was added. Both these models yield a value \HeII/\HI=35 
in optically thin absorbers. In the HM+S case the 
effective \Lyb\ line opacity is 0.43, and the sawtooth modulation causes a small decrease in the 
metagalactic flux, by at most a factor of 2, relative to the old HM spectrum.          
We have also run 3 other representative cases, termed ``DR" for
``delayed reionization". These models ignore the patchy nature of the reionization process and assume that
a larger fraction of intergalactic helium at redshift 3 is in \HeII: this is obtained by 
artificially increasing the \HeII/\HI\ ratios computed by CUBA to 160, 250, and 530, 
respectively. Resonant absorption from the Lyman series now causes a reduction of the 
background intensity between 3 and 4 Ryd by as much as one dex (off-resonance) compared to the HM spectrum.  

\section{Discussion}

Since the pioneering work of Chaffee \etal (1986) and Bergeron \& Stasinska (1986), many studies have
used observations of intervening metal absorption systems to reconstruct the shape of the photoionizing 
radiation field at $z\lta 3$. The many modifications to the HM96 background intensity that have been 
proposed include: 1) a stronger \HeII\ \Lya\ feature in order to match the observed \SiIV/\SiII\ 
abundance ratios (Levshakov \etal 2003); 2) a depression between 3 and 4 Ryd in order to enhance 
the predicted \CII/\CIV\ and \CIII/\CIV\ ratios, incorrectly attributed by Agafonova \etal (2007) 
to a \HeII\ \Lya\ Gunn-Peterson effect; 3) a softer far-UV spectrum in order to predict [O/Si] 
values that are consistent with theoretical yields (Aguirre \etal 2008). A detailed modelling 
of the abundances of intergalactic metals is beyond the scope of this paper: here, we just want to point 
out that the sawtooth modulation, if as large as computed in the DR models, may provide a better match to 
the observations. At the top of the right panel of Figure \ref{fig1} we have indicated the positions 
of the ionization thresholds of \SiIII\ (33.5 eV), \SiIV\ (45.1 eV), \CIII\ (47.9 eV), \CIV\ (64.5 eV), 
\OII\ (35.1 eV), and \OIII\ (54.9 eV) ions. The reprocessing of Lyman series and Lyman continuum photons
increases \HeII\ \Lya\ in the DR spectra by a factor of 1.7 compared to the HM case. 
The flux at the \SiIV\ ionization threshold decreases by a factor of 2, boosting the 
predicted abundance of \SiIV. An even larger boost is expected in the abundance of \CIII, 
whose ionization threshold lies exactly within the \HeII\ \Lyb\ deep absorption feature, and 
of \OIII, whose threshold lies just beyond the \HeII\ Lyman limit.  

The above results show that line absorption from the Lyman series of intergalactic helium 
may be an important, so far neglected, process shaping the spectrum of the cosmic radiation 
background above 3 Ryd. The large resonant cross-sections for far-UV light scattering make the 
sawtooth modulation a sensitive probe of the epoch of helium reionization and of the sources 
that keep the IGM highly ionized. The \HeII\ sawtooth may be one of the crucial missing ingredients 
in the modelling of the abundances of metal ions like \CIII\ and \SiIV\ observed in the IGM at redshift 3. 
In the case of large line opacities, substantial fluctuations are expected in the far-UV background intensity 
near each resonance, as the first integral in equation (\ref{Jbeta}) extends over a small 
absorption distance, and just a few quasars are expected to contribute to the local emissivity. 
Such fluctuations may cause large variations in e.g. the observed \CIII\ abundances. 
In future work, we intend to study such fluctuations and address how the contribution of star-forming 
galaxies to the background may affect the \HeII/\HI\ ratio in the IGM and the predicted sawtooth modulation. 

\acknowledgments
Support for this work was provided by NASA through grants HST-AR-11268.01-A1
and NNX08AV68G (P.M.). We thank Steve Furlanetto, Zoltan Haiman, and Avery Meiksin for 
useful discussions. The spectra plotted in Figure \ref{fig1} are available upon request.

{}

\end{document}